\documentclass{ere04}

\usepackage{graphicx}    

\usepackage{amsfonts}                    
\usepackage{amssymb}  

\def\beq{\begin{equation}}
\def\eeq{\end{equation}}

\def\bea{\begin{eqnarray}}
\def\eea{\end{eqnarray}}
\def\ba{\begin{array}}                  
\def\ea{\end{array}}

\def\makeatletter{\catcode`\@=11}
\makeatletter
\def\mathbox#1{\hbox{$\m@th#1$}}%
\def\math@ccstyles#1#2#3#4#5#6#7{{\leavevmode
      \setbox0\mathbox{#6#7}%
      \setbox2\mathbox{#4#5}%
      \dimen@ #3%
      \baselineskip\z@\lineskiplimit#1\lineskip\z@
      \vbox{\ialign{##\crcr
             \hfil \kern #2\box2 \hfil\crcr
             \noalign{\kern\dimen@}%
             \hfil\box0\hfil\crcr}}}}
\def\mathaccstyles{\math@ccstyles\maxdimen}
\def\maththroughstyles{\math@ccstyles{-\maxdimen}}
\def\unity%
 {\maththroughstyles{.45\ht0}\z@\displaystyle {\mathchar"006C}\displaystyle 1}
\begin{document}

\title*{Giant Gravitons as Fuzzy Manifolds}

\titlerunning{Giant Gravitons as Fuzzy Manifolds}


\author{B. Janssen \inst{1}, Y. Lozano\inst{2} \and
D. Rodr\'{\i}guez-G\'omez\inst{2}}


\institute{Dpto de F\'{\i}sica Te\'orica y del Cosmos and C.A.F.P.E.,
           Univ. de Granada, 18071 Granada, Spain. \texttt{bjanssen@ugr.es}
\and Departamento de F\'{\i}sica, Universidad de Oviedo, Avda. Calvo 
     Sotelo 18, 33007 Oviedo, Spain. 
\texttt{yolanda@string1.ciencias.uniovi.es}, 
\texttt{diego@fisi35.ciencias.uniovi.es}}
%
%
\maketitle

Giant gravitons are described microscopically in terms of dielectric gravitational
waves expanding into fuzzy manifolds.
We review these constructions in $AdS_m\times S^n$ spacetimes, discussing the
different fuzzy manifolds that appear in each case.\footnote{Talk
given by D.R.G. at the XXVII Spanish Relativity Meeting in Madrid, September 2004.}

\section{Introduction}
\label{sec:1}

Giant gravitons \cite{GST}-\cite{HHI} are stable brane configurations
with non-zero angular momentum, that are wrapped around $(n-2)$- or
$(m-2)$-spheres in $AdS_m \times S^n$ spacetimes, and carry dipole moment 
with respect to the background gauge potential. They are 
stable because the contraction due to
the tension of the brane is precisely cancelled by the expansion due to
the coupling of the angular momentum to the background flux.  These spherical 
brane configurations turn out to be massless, conserve
the same number of supersymmetries and carry the same quantum numbers of a
graviton. The fact that they are extended objects of finite size has lead
to the name of giant gravitons.

These configurations were first proposed in \cite{GST} as a way to satisfy
the stringy exclusion principle implied by the $AdS/CFT$ correspondence
\cite{EP}. The spherical $(n-2)$-brane expands into the $S^n$ part of the
geometry with a radius proportional to its angular momentum. Since this
radius is bounded by the radius of the $S^n$, the configuration has
associated a maximum angular momentum. There are as well \cite{GMT,HHI}
``dual'' giant graviton configurations, corresponding to spherical
$(m-2)$-branes expanding in the $AdS_m$ part of the spacetime, and thus
not satisfying the stringy exclusion principle. Possible ways of realizing
this principle in the presence of these degenerate solutions have been
proposed in for example \cite{GMT,BS}.

The appearance of giant gravitons as blown up massless particles hints to
a connection with other examples of expanded brane configurations, more
precisely to the dielectric effect \cite{Myers}, where multiple coinciding D$p$-branes
can expand into higher dimensional D-brane configurations. There are two
complementary descriptions of this effect. Consider the case in which the D$p$-branes
expand into a spherical D$(p+2)$-brane.
The first one is an
Abelian, macroscopic description, describing the spherical D$(p+2)$-brane with 
a large number of D$p$-branes dissolved on its worldvolume \cite{roberto}. The
second one is a non-Abelian, microscopic formulation \cite{TvR1}-\cite{Myers}, 
describing how multiple coinciding D$p$-branes expand into a D$(p+2)$-brane with 
the topology of a fuzzy 2-sphere \cite{Myers}. Both descriptions agree in the
limit where the number $N$ of D$p$-branes is very large.

It has been suggested in the literature \cite{DTV}-\cite{JL1} that there 
should exist a microscopic description of the Abelian giant
gravitons of \cite{GST}-\cite{HHI} in terms of dielectric gravitational
waves. Since massless particles, in particular gravitons, are the source
terms for gravitational waves, it is natural to expect that a dielectric
effect for gravitational waves will provide a microscopic picture
for the giant graviton configurations. By analogy with the dielectric
effect for D-branes, it is believed that in the limit when the number of
gravitons is large, this microscopic description should match the macroscopical 
description of \cite{GST}-\cite{HHI}, where now the angular momentum of the 
giant graviton is interpreted  microscopically as the total momentum 
of the multiple gravitational waves.

The actions describing coincident gravitational waves in 
Type IIA and M-theory were derived in \cite{JL1,JL2},
using Matrix string theory and a chain of 
dualities. These actions contain the couplings to closed string backgrounds, and
in particular the multipole couplings that give rise to the dielectric effect.
With these actions, it is then possible to check explicitly the 
claims of \cite{DTV}-\cite{JL1} about the microscopical description of giant 
gravitons. This has been done in a series of papers \cite{JL2}-\cite{JLR2} for 
the different $AdS_m \times S^n$ backgrounds. The aim of this paper is to review 
these constructions, emphasising the specific details of each case.
In particular we will 
look at the $AdS_m \times S^n$ backgrounds relevant for string theory, namely 
$AdS_7 \times S^4$, $AdS_5 \times S^5$, $AdS_4 \times S^7$ and 
$AdS_3 \times S^3 \times T^4$. As the construction of the dual giant gravitons
in each case is very similar to the genuine ones, we will not deal with the former
here but refer to \cite{JL2}-\cite{JLR2}.

\section{Macroscopic giant gravitons}
   
In suitable coordinates the $AdS_m\times S^n$ backgrounds can be written as
\bea
&&ds^2= -\big(1+{\textstyle{\frac{r^2}{\tilde{L}^2}}}\big)dt^2
      + \big(1+{\textstyle{\frac{r^2}{\tilde{L}^2}}}\big)^{-1}dr^2 + r^2d\Omega_{m-2}^2
      + L^2\big(d\theta^2+\cos^2\theta d\phi^2+\sin^2\theta  d\Omega_{n-2}^2\big), 
\nonumber \\
&&C_{\phi\chi_1\cdots\chi_{n-2}}^{(n-1)}=
\beta_nL^{n-1}\sin^{n-1}\theta\sqrt{g_{\chi}}, 
\hspace{2cm}
C_{t\alpha_1\cdots\alpha_{m-2}}^{(m-1)}=
-{r^{m-1}}\tilde{L}^{-1}\sqrt{g_{\alpha}}
\label{background}
\eea
where $L=\frac{n-3}{2}\tilde{L}$ and $\beta_4=\beta_5=-\beta_7=1$. $d\Omega_{n-2}$ 
($d\Omega_{m-2}$) and $\sqrt{g_{\chi}}$ ($\sqrt{g_{\alpha}}$) stand respectively for 
the metric and the volume form of a unit $(n-2)$-sphere ($(m-2)$-sphere)
parametrized with the coordinates $\{\chi_i\}$ ($\{\alpha_i\}$). The 
$AdS_3\times S^3\times T^4$ background is special. It arises as the near horizon
geometry of the intersecting D1-D5 system:
\bea
\label{ads3}
&&\hspace*{-.3cm}
ds^2=-\big(1+{\textstyle{\frac{r^2}{\tilde L^2}}}\big)dt^2
      +\big(1+{\textstyle{\frac{r^2}{\tilde{L}^2}}}\big)^{-1}dr^2 + r^2d\varphi^2
      + L^2\big(d\theta^2+\cos^2\theta d\phi^2+\sin^2\theta d\chi^2\big) + R^2dy_a^2,  
\nonumber\\
&& \hspace*{1cm}
e^{\Phi}=R^2, \hspace{2cm}
C^{(2)}_{t\varphi}=-Q_5L^{-3}r^2, \hspace{2cm}
C^{(2)}_{\varphi\chi}=Q_5\sin^2\theta, 
\eea
where $Q_1$ and $Q_5$ are the total D1- and D5-brane charges, and
$y_a$ ($a=1,\dots ,4$) describe the relative transverse space.
$R$, $L$ and $Q_5$ are related via $L^2 = Q_5 R^2$. 

Consider a test $(n-2)$-brane wrapped around the $S^{n-2}$ of (\ref{background}) and  
moving along the $\phi$ direction (i.e. take the Ansatz $\theta =$ constant, $r=0$, 
$\phi=\phi(t)$). Its energy is easily seen to be given by \cite{GST}
\begin{equation}
\label{potmacro}
E=\frac{P_\phi}{L}\sqrt{ 1 + \tan^2\theta 
            \left( 1 -{\frac{\tilde N}{P_\phi}}\sin^{n-3}\theta \right)^2}.
\end{equation}
Here $\tilde N = T_{(n-2)} A_{(n-2)} L^{n-1}$ is an integer coming from the flux 
quantization on the $S^{n-2}$ (with $T_p$ the tension of the $p$-brane and $A_p$ 
its area) and $P_\phi$ is the angular momentum in $\phi$. Minimizing 
with respect to $\theta$ one finds two minima, both with energy $E =P_\phi/L$, 
namely $\sin\theta=0$ and $\sin\theta=(P_\phi/{\tilde N})^{\frac{1}{n-3}}$. Since 
the radius 
of the $(n-2)$-sphere in (\ref{background}) is given by $L\sin\theta$, it is 
clear that the first solution corresponds to a point-like particle, while the
second one is the so called giant graviton, an expanded brane of finite size,  
whose quantum numbers are in fact those of a massless particle \cite{GST,GMT}.
Notice that since $\sin\theta$ 
is bounded,  $P_\phi$ is also bounded, thus leading to a realisation of the string 
exclusion principle \cite{GST}. 

In the $AdS_3\times S^3 \times T^4$ background the most general giant graviton solution
is in terms of a test brane consisting on a bound state of D1-branes and D5-branes
wrapped on the 4-torus. Furthermore, since the 
cycles in $AdS_3$ and $S^3$ are both $S^1$, it is possible to construct mixed giant 
gravitons, a linear combination of genuine and dual ones. Here we will only deal with 
the simplest case of a D1-brane wrapped around the $S^1\subset S^3$. For details of 
the general 
construction we refer to \cite{LMS}. The energy of this brane is given by 
\beq
E=\frac{P_\phi}{L}\sqrt{ 1 + \tan^2\theta 
                  \left( 1 -{\frac{\tilde N}{R^2P_\phi}} \right)^2}.
\label{Eads3}
\eeq
The minimum energy is reached when $\theta =0$ or $P_\phi= {\tilde N}/R^2$. Note that the 
giant graviton solution does not put constraints on the radius of the brane and that 
this solution only exists for specific values of the momentum. This poses a puzzle with
the realisation of the stringy exclusion principle (see \cite{HHI,LMS}).

\section{The action for M-theory gravitational waves}
 \label{sec:2}

The action for $N$ 11-dim gravitational waves in an arbitrary background is 
given by \cite{JL2}
\begin{eqnarray}
S&=&-T_W \int d\tau \ \mbox{STr} 
    \left\{ k^{-1}\sqrt{-P[E_{00}+E_{0i}(Q^{-1}-\delta)^i_k E^{kj}E_{j0}]\ 
                    \det Q} \ \right\}
\nonumber
\\ 
&+&T_W \int d\tau \ \mbox{STr} \left\{ -P[k^{-2} k^{(1)}]
                    +i P[(\mbox{i}_X \mbox{i}_X)C^{(3)}] +
                    \textstyle{\frac12} P[(\mbox{i}_X \mbox{i}_X)^2 \mbox{i}_k C^{(6)}] 
                    +\cdots  \right\}, 
\label{11daction}\\
&& \hspace*{-.8cm} 
E_{\mu\nu}= g_{\mu\nu}- k^{-2}k_\mu k_\nu +k^{-1}(\mbox{i}_k C^{(3)})_{\mu\nu}, \nonumber
\hspace{2cm}
Q^i_j=\delta^i_j + ik[X^i,X^k]E_{kj}.
\end{eqnarray}
Here $T_W$ is the (momentum) charge of a single graviton. This action contains the
direction of propagation of the waves as a special isometric direction, with
Killing vector $k^\mu$. In the Abelian limit, a Legendre transformation restoring the
dependence on this direction yields the usual action for massless particles. In turn,
in the non-Abelian case, this action gives rise to Myers action for coincident D0-branes
after dimensional reduction over $k^\mu$, and the action for coincident Type IIA
gravitational waves, obtained in \cite{JL1} via Matrix string theory in a weakly curved
background, after dimensional reduction along a different direction. Consistently,
the Matrix theory
calculation yields as well an isometric action for coincident Type IIA waves.
Notice that the waves are minimally coupled to the momentum operator
$k^{(1)}_\mu/k^2=g_{z\mu}/g_{zz}$, in coordinates adapted to the isometry in
which $k^\mu=\delta^\mu_z$.

\section{The microscopic description for giant gravitons}

Using the action (\ref{11daction}) we now provide
a microscopical description for the giant gravitons in $AdS_m\times S^n$ spacetimes in terms 
of expanding gravitational waves.

\subsection{Giant gravitons in $AdS_7\times S^4$: the standard case}

{}From the Abelian picture we know that the giant gravitons should come from
waves propagating in the $\phi$-direction and expanding into 
a non-Abelian version of the $S^2$. We thus identify the isometry direction of 
the action (\ref{11daction}) as $k^\mu = \delta^\mu_\phi$ and parametrise the 
$S^2$ in terms of cartesian coordinates $X^i=\frac{L\sin\theta}{\sqrt{N^2-1}}J^i$, 
with $J^i$ forming a $N$-dim representation of $SU(2)$. Since the $X^i$ satisfy 
$\sum_{i}(X^i)^2=L^2\sin^2\theta \unity$, through the quadratic Casimir of the group,
they span a fuzzy 
two-sphere. Substituting this Ansatz in the action  yields an energy \cite{JL2}:
\begin{equation}
E=\frac{T_W}{L\cos\theta}\mbox{STr}\left\{ 
           \sqrt{\unity -{\frac{4L\sin\theta}{\sqrt{N^2-1}}}X^2
          + {\frac{4L^4\sin^2\theta\cos^2\theta}{N^2-1}}X^2
          + {\frac{4L^2\sin^2\theta}{N^2-1}} X^2 X^2} \ \right\}.
\label{action7-4}
\end{equation}
Given that we are interested in the large $N$ limit in order to compare with the 
Abelian calculation, and taking into account that 
$\mbox{STr}\{(X^2)^n\}=\mbox{Tr}\{(X^2)^n\} + \mathcal{O}(\frac{1}{N^2-1})$,
we can rewrite the action (\ref{action7-4}) for large $N$ as
\begin{equation}
E=\frac{NT_W}{L}\sqrt{1+\tan^2\theta
                \Bigl(1-{\frac{2L^3}{\sqrt{N^2-1}}}\sin\theta\Bigr)^2}.
\end{equation}
Identifying the macroscopic momentum $P_\phi$ with the sum of the charges of the $N$ 
gravitons and taking into account that ${\tilde N}=T_2 A_2 L^3$
we find perfect agreement upto order $N^{-2}$ with the macroscopical
computation (\ref{potmacro}). We also find agreement to the same 
order for the point-like and giant graviton solutions and the upper bound on the momentum.

\subsection{Giant gravitons in $AdS_5\times S^5$: the Hopf fibration}   
\label{sec:3}

The action for coincident gravitational waves in Type IIB can be derived via 
reduction plus T-duality 
from the action (\ref{11daction}). The BI part is of the same form 
as in (\ref{11daction}), where now 
\bea
E_{\mu\nu} = g_{\mu\nu}-k^{-2}k_\mu k_\nu -l^{-2}l_\mu l_\nu
-k^{-1}l^{-1} e^\Phi (\mbox{i}_k\mbox{i}_l C^{(4)})_{\mu\nu}, \hspace{.5cm}
Q^i_j = \delta^i_j + i[X^i,X^k] e^{-\Phi} k l E_{kj}.
\nonumber
\eea
Here $l^\mu$ is a second Killing vector, pointing along the direction in which 
the T-duality is performed, which appears explicitly in the action for coincident
Type IIB waves (see \cite{JLR1}).
In the CS part, the coupling to the RR 4-form is of the 
form $P[(\mbox{i}_X \mbox{i}_X) \mbox{i}_l C^{(4)}]$.

The construction of non-Abelian giant gravitons in $AdS_5 \times S^5$ turns out 
to be more involved than the ones in $AdS_7 \times S^4$. Here 
the gravitational waves expand into a spherical D3-brane, and hence a fuzzy 
3-sphere Ansatz is needed.
Yet the extra isometry in the Type IIB action is of help. The presence of this
compact isometry suggests the representation of the 3-sphere as a $U(1)$ bundle over
$S^2$, with the $U(1)$ invariance associated to translations along this direction.
The fuzzy
$S^3$ is then constructed as an Abelian $U(1)$ fibre over a fuzzy $S^2$.  The 
details of the construction can be found in \cite{JLR1}.

Parametrising the $S^2$ base manifold as in 4.1 by the cartesian coordinates
$X^i =\frac{\tilde R}{\sqrt{N^2-1}}J^i$, with $\tilde R$ the radius of the $S^2$ and 
$\sum_i (X^i)^2 =\tilde R^2 \unity$, 
we find that the energy is given by \cite{JLR1}
\begin{equation}
\hspace*{-.3cm}
E = {\frac{T_W}{L\cos{\theta}}} {\rm STr} \left\{ 
\sqrt{\unity - {\frac{L^4\sin^4{\theta}}{2 {\tilde R}^2\sqrt{N^2-1}}} X^2
       + {\frac{L^8\sin^6{\theta}\cos^2{\theta}}{16{\tilde R}^2(N^2-1)}}X^2
       + {\frac{L^8\sin^8{\theta}}{16{\tilde R}^4(N^2-1)}} (X^2)^2}\right\} ,
\end{equation}
which for large $N$ can again be rewritten to a form that agrees with the Abelian case
(\ref{potmacro}):
\begin{equation}
\label{potmic}
E={{\frac{NT_0}{L}}}\sqrt{1+\tan^2{\theta}
\Bigl(1-{\frac{L^4\sin^2{\theta}}{4\sqrt{N^2-1}}}\Bigr)^2}.
\end{equation}

\subsection{Giant gravitons in $AdS_4\times S^7$: fuzzy $CP^2$}   

The giant graviton in this spacetime is an $S^5$, which can be described as a 
$U(1)$-bundle over
the two dimensional complex projective plane, $CP^2$. It is again suggestive to 
identify the isometric direction in the action with the $U(1)$ fibre. However,
the action for M-theory gravitational waves contains only one isometric direction, which
is typically identified with the direction of
propagation of the
waves. The relevant dielectric coupling in (\ref{11daction}) to the
6-form potential is
$P[(\mbox{i}_X \mbox{i}_X)^2 \mbox{i}_k C^{(6)}]$,  which is however vanishing
for the 6-form potential of the $AdS_4\times S^7$ background, given by 
(\ref{background}). This is however not the case if we identify 
$k^\mu$ with the coordinate along the
$U(1)$ fibre in the Hopf fibration of the $S^5$, and 
the propagation
direction $\phi$ is in turn taken to be an isometry which is not explicit 
in the action (\ref{11daction}).  
Analogously to the previous case, the
fuzzy $S^5$ onto which the gravitons expand is then the Hopf fibre of
the fuzzy $CP^2$,   
whose construction is known in the literature (see for instance \cite{ABIY}).
Taking into account that $CP^2$ can be defined as a submanifold
of ${\mathbb{R}}^8$ determined by certain constraints, its fuzzy version can be obtained
impossing these conditions at the level of matrices. This is satisfied for
$X^i=\frac{1}{\sqrt{n^2+3n}}T^i$, where $T^i$ are the
$SU(3)$ generators in the $(n,0)$ or $(0,n)$ irreducible representations.
(For the technical details, we refer to \cite{JLR3}.)  
Taking into account that the dimension of these representations is given by 
$N= (n+1)(n+2)/2$,
the Lagrangian of the system of waves is, upto order $N^{-2}$,
\begin{equation}
\label{lagcp2}
{\cal L}=-\frac{T_W}{L\sin\theta} \mbox{STr}
     \left\{ \sqrt{1-L^2\cos^2{\theta}{\dot{\phi}}^2}
              \Bigl( \unity + {\frac{3L^6 \sin^6 \theta}{8(N-1)}} X^2 \Bigr)
          \  - \ {\frac{9 L^7 \sin^7\theta}{8(N-1)}}\dot{\phi}(X^2)^2 
 \right\}. 
\end{equation} 
Eliminating the velocity $\dot{\phi}$ 
via a Legendre transformation, we obtain for the energy \cite{JLR3}
\begin{equation}
\label{Hgiant}
E=\frac{P_\phi}{L}\sqrt{1+\tan^2{\theta}
          \Bigl(1-{\frac{NT_0}{8(N-1)P_\phi}} L^6\sin^4{\theta}\Bigr)^2
     +{\frac{N^2T_0^2}{P_\phi^2\sin^2{\theta}}}
           \Bigl(1+{\frac{L^6\sin^6{\theta}}{4(N-1)}}\Bigr)}.
\end{equation}
Here we have to recall that the gravitons described by the Lagrangian 
(\ref{lagcp2}) carry, by construction, momentum along the $U(1)$ fibre
isometric direction,  that we parametrize by $\chi$, 
this momentum being given by $P_\chi=NT_W$.
Therefore, in order to make contact with the Abelian giant graviton of
(\ref{potmacro}) we have to switch to zero this momentum, without setting to
zero the number of expanding gravitons!
The difference between $P_\chi$ being zero or not is merely a coordinate
transformation, a boost in $\chi$. How to perform coordinate transformations
in non-Abelian actions is however an open problem. Still, we can compare the
gravitons described by the Lagrangian (\ref{lagcp2}) with the corresponding 
macroscopical description, which is in terms of a spherical M5-brane moving in
both the $\phi$ and $\chi$ directions.  In this description the momentum along
$\chi$ can be neatly set to zero, since an M5-brane moving only along the
$\phi$ direction is perfectly well-defined. This shows that, at least for large $N$,
setting $P_\chi$ to zero eliminates the last term in (\ref{Hgiant}).
We then obtain an expression that 
agrees with (\ref{potmacro}) for infinite number of gravitons:
\begin{equation}
\label{Hmic}
E=\frac{P_\phi}{L}\sqrt{1+\tan^2{\theta}
               \Bigl(1-{\frac{NT_0}{8(N-1)P_\phi}}L^6\sin^4{\theta}\Bigr)^2}.
\end{equation}

\subsection{Giant gravitons in $AdS_3\times S^3\times T^4$: the worldvolume scalar}   

Giant gravitons in $AdS_3\times S^3\times T^4$ are branes wrapped around an 
$S^1$, which raises the immediate question of how to construct a non-Abelian 
realisation, as the $S^1$ has associated
an (Abelian) $U(1)$ algebra. The solution is to embed the $S^1$ in a cylinder, which
can then be made fuzzy. In order to do this we have to take the worldvolume scalar
associated by T-duality with the duality direction non-vanishing. This field will
not have a geometrical meaning, since it is not a transverse scalar, but plays the role
of the direction along the axis of the cylinder in the construction of the algebra.

The BI part is again of the form (\ref{11daction}), but now we have to consider as
well the contribution of the non-vanishing worldvolume scalar, $\omega$,
(see \cite{JLR2} for the
explicit expresion). The relevant CS coupling is now 
${\rm STr}\{ [X^i,\omega] C^{(2)}_{ij}DX^j \}$.

%
%

Considering a fuzzy cylinder whose spatial section is a 
circle described by cartesian coordinates $X^i$ ($i=1,2$) and whose axis is 
along the scalar 
field $\omega$, the algebra is given by:
\begin{equation}
\label{fuzzycyl}
[X^i,X^j]=0, \hspace{2cm}
[X^i,\omega]= i \varepsilon^{ij} f X^j,
\end{equation}
with $f$ the non-commutative parameter. The quadratic Casimir gives 
$\sum_i (X^i)^2 = L^2 \sin^2\theta \unity$. All the representations of this algebra 
are however infinite dimensional, so we are forced to deal with an infinite 
number of gravitons. Substituting in the action we find an energy \cite{JLR2}:
\bea
E= \frac{T_W}{L \cos\theta} \mbox{STr}\left\{
     \sqrt{\Bigl(\unity -{\frac{fQ_5}{L^2}}X^2\Bigr)^2 
              + {\frac{f^2 Q_5^2}{L^2}}\cos^2{\theta} X^2 } \  
\right\}.
\eea
The energy per unit length of the cylinder is then given by
\begin{eqnarray}
\label{micpot2}
E&=&\frac{T_W}{fL} \sqrt{1 + \tan^2 \theta ( 1 - f Q_5)^2},
\end{eqnarray}
where we have taken into account that the length of the cylinder can be estimated
as $l=f {\rm Tr}\unity$.
As in the macroscopic case, we find giant gravitons of arbitrary
radius if $f = Q_5^{-1}$. Taking into account that the
microscopical momentum per unit length is given by ${\rm Tr}{\unity}T_W/l$, which
is what should be compared with $P_\phi$ in (\ref{Eads3}), we find perfect
agreement between the two computations. This is in agreement with the fact that
we need an infinite number of gravitons in order to span the fuzzy cylinder.

\section*{Acknowledgements}

It is a pleasure to thank the organizers of the XXVII Spanish Relativity
Meeting. The work of B.J. is done as part of the program ``Ram\'on y Cajal''
of the M.E.C. (Spain). He was also partially supported by the M.E.C. under
contract FIS 2004-06823 and by the Junta de Andaluc\'{\i}a group FQM 101.  
The work of Y.L. and D.R-G. has been partially supported by CICYT grant
BFM2003-00313 (Spain).  D.R-G. was supported in part by a F.P.U.
Fellowship from M.E.C.

 
\end{document}